\documentclass[12pt]{iopart}
\usepackage{graphicx}
\input epsf.tex
\begin{document}
\title[Identified particle $v_{2} (\eta ,p_{t})$ at RHIC]{Rapidity and $p_{t}$ 
dependence of identified-particle elliptic flow at RHIC}
\author{S J Sanders for the BRAHMS Collaboration\footnote{For the full
authorlist and acknowledgements see the appendix of this volume.}}
\address{The University of Kansas, Lawrence, KS  66045, USA}
\ead{ssanders@ku.edu}	
\begin{abstract}
Elliptic flow has been measured by the BRAHMS experiment as a function
of transverse momentum and pseudorapidity for the Au+Au reaction at
$\sqrt{s_{NN}} = 200~GeV$.   Identified-particle 
$v_{2} (\eta ,p_{t})$
values were obtained with the two BRAHMS spectrometers at pseudorapidities
$\eta \approx$ 0, 1, and 3.4.   The results show that the differential 
$v_{2} (\eta ,p_{t})$ values for a given particle type are 
essentially constant over the
covered pseudorapidity range. It is suggested that the dominant cause of 
the observed
fall-off of the integral $v_{2}$ values going away
from mid-rapidity is a corresponding softening 
of the particle spectra .   
\end{abstract}	
\submitto{\JPG}	

The large azimuthal anisotropy in particle production observed near 
mid-rapidity at RHIC has been taken as evidence for the formation 
of an almost ideal fluid corresponding to a  strongly interacting 
quark-gluon plasma\cite{Shuryak:2003xe}.
Pressure gradients of the thermalized  medium formed 
early in the relativistic heavy-ion collisions result in azimuthally 
asymmetric particle production, with the highest final particle densities 
occurring in the reaction plane. 
Elliptic flow is identified with the $v_{2}$, $2^\mathrm{nd}$-harmonic term 
of a Fourier expansion of the azimuthal angular dependence of the particle
production. Near mid-rapidity, the 
identified particle $v_{2} (p_{t})$ values are found to be 
near the hydrodynamic limit,
with a particle mass dependence below 2 GeV close to that 
expected from hydrodynamic models\cite{Adams:2004bi,Adler:2003kt}. 
The $v_{2}$ values integrated over 
transverse momentum are found to  fall 
off going away from mid-rapidity, dropping by
about 35\% by pseudorapidity $\eta = 3$, with similar behavior seen in 
Au+Au reactions from $\sqrt{s_{NN}}$ = 19.6 GeV to 200 GeV\cite{Back:2004zg}.
   
As part of its program to determine how the physics at RHIC changes in
going to forward rapidities\cite{Arsene:2004fa}, BRAHMS has measured
$v_{2}$ as a function of transverse momentum for pions,
kaons, and protons at angles (pseudorapidities)
$90^\mathrm{o} (\eta = 0)$, $40^\mathrm{o} (\eta = 1)$,
 and $4^\mathrm{o} (\eta = 3.4)$. The experiment
has previously presented evidence that the charged-particle and
pion  $v_{2} (p_{t})$ behavior
at forward rapidities differs little from that seen near mid-rapidity,
a somewhat surprising finding in light of the integral $v_{2}$ 
results\cite{Ito:2006dm}. The new analysis explores this behavior further by 
studying additional particle channels and by analyzing the 
$v_{2}(p_{t})$ behavior together with the associated particle
spectra.   

The orientation of the reaction plane with respect to the laboratory
system was determined using azimuthally symmetric rings of 
Si strip detectors and  scintillator tile 
detectors in the BRAHMS Multiplicity Array, and plastic Cherenkov
radiators mounted to phototubes in one of the experiment's Beam-Beam counter 
arrays\cite{Adamczyk:2003sq}. The general procedure for determining
the reaction plane and the corresponding reaction-plane resolution
is described by Poskanzer and Voloshin\cite{Poskanzer:1998yz}. The
reaction plane corresponding to the second moment of the angular
distribution is found with
\begin{equation}
\Psi _2  = {1 \over 2}\sum\limits_i {{{w_i \sin \left( {2\phi _i } \right)} 
\over {w_i \cos \left( {2\phi _i } \right)}}} ,
\end{equation}
where the sum is over all detector elements with geometric weighting
factors $w_{i}$ and azimuthal angle $\phi _{i}$. 

The angular correlation with respect to the
reaction plane of particles detected in the 
BRAHMS mid-rapidity and forward spectrometers  determine the observed $v_{2}$
values, with
\begin{equation}
v_2^{obs}  = \left\langle {\cos \left( {2\left[ {\phi  - \Psi _2 } \right]} \right)} \right\rangle .
\end{equation} 
Here the angular bracket denotes an average over all events of a given class,
such as all pions within a given range of transverse momenta. 
Since the BRAHMS spectrometers are small acceptance devices,
all particle detected in the mid-rapidity spectrometer have 
$\phi \approx 0^{o}$, while those detected by the forward spectrometer have 
$\phi \approx 180^{o}$.
 
The true $v_{2}$ value, which is based on the actual reaction plane,
is found from the $v_{2}^{obs}$ value using the reaction-plane correction
factor R, with
\begin{equation}
v_2  = v_2^{obs} /R  .
\end{equation}
The reaction plane correction for any given
ring (or ring combination) is determined using two additional 
rings that
are far enough apart so as to avoid autocorrelations.  Considering
rings A, B, and C, the reaction-plane correction factor for ring A
is found as:
\begin{equation}
{\rm{R}}_{\rm{A}} {\rm{ = }}\sqrt {{{\left\langle {\cos \left( 
{2\left[ {\Psi _2^A  - \Psi _2^B } \right]} \right)} \right\rangle 
\left\langle {\cos \left( {2\left[ {\Psi _2^A  - \Psi _2^C } \right]} 
\right)} \right\rangle } \over {\left\langle {\cos \left( {2\left[ 
{\Psi _2^B  - \Psi _2^C } \right]} \right)} \right\rangle }}} 
\end{equation}
The reaction plane resolution factor was typically in the range
$0.18 < R < 0.25$.

\begin{figure}[htbp]	
\begin{center}	
\includegraphics[width=135mm]{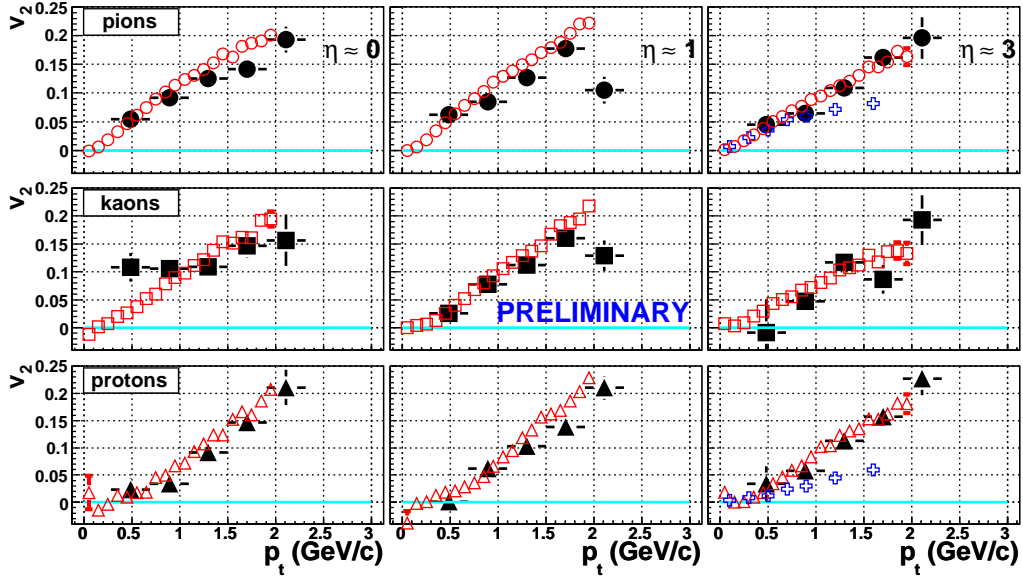}
\caption{
$\mathrm{v_{2}(p_{t}, \eta )}$ for pions, kaons, and protons (filled symbols).
Theoretical values based on the hydrodynamic calculations of Hirano 
{\it et al.}~\cite{Hirano:2005xf} are shown by the open circles, squares, and triangles.  The
open crosses correspond to  $\mathrm{\eta \approx 4}$ calculations using
the AMPT model~\cite{Chen:2004vh}. 
\label{fig:fig1}
}
\end{center}
\end{figure}

Figure~\ref{fig:fig1} shows the resulting values (closed symbols) 
of $v_{2}(p_{t})$ for pions, kaons, and protons at the indicated pseudorapidities, selecting events in the 10\% - 50\% centrality class. 
The behavior for the three different particle species shows very little
change with pseudorapidity.

\begin{figure}[htbp]	
\begin{center}	
\includegraphics[width=80mm]{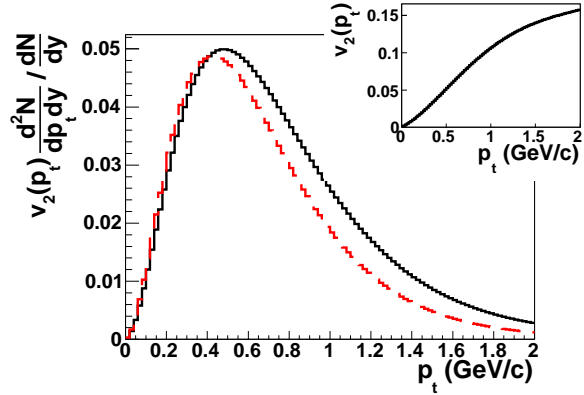}
\caption{
Illustration of how the softening of the particle spectra going to forward
angles affects the integral $\mathrm{v_{2}}$ values.  The insert shows the
general behavior of the differential $\mathrm{v_{2}(p_{t})}$ values for pions,
which for this illustration is assumed to be independent of 
pseudorapidity.  Folding this distribution with the corresponding 
experimentally observed pion
spectra at $4^{o}$ and $90^{o}$ results in the dashed and solid curves, 
with integral $\mathrm{v_{2}}$ values of 0.046 and 0.036, respectively.
\label{fig:fig2}
}
\end{center}
\end{figure}

The average value of transverse momentum for pions in the
10\% - 50\% centrality class at $90^\mathrm{o}$ is
($475\pm 60$)~MeV/c, falling to ($380\pm 45$)~MeV/c at $4^\mathrm{o}$. These
values are obtained by fitting a power-law dependence to the respective
particle spectra.
The softening of the particle spectrum going to more forward rapidities 
can have a significant effect on the  integral $v_{2}$ values.  
This is illustrated
in Figure~\ref{fig:fig2}. The insert shows an assumed form of $v_{2}(p_{t})$ 
that is
taken as being constant as a function of pseudorapidity.  
Folding this behavior with
the experimentally observed normalized particle spectra for pions at 
$90^\mathrm{o}$ 
and $4^\mathrm{o}$ result in the
solid and dashed lines of the main figure panel, respectively. The softer spectrum 
at the more forward
pseudorapidity results in the weighted $v_{2}$ distribution peaking at a lower
mean $p_{t}$ value, resulting in a 22\% smaller integral $v_{2}$ 
value than a mid-rapidity. 
Two-thirds of the observed integral $v_{2}$ change is then 
attributable to the softening of the particle spectrum.     

The small change in the differential $v_{2}$ signal going from mid- to forward
rapidity  suggests a longitudinally extended region for the medium 
produced in the collision.  Rapidity dependent changes  in the 
radial flow or other rescattering behavior of the
hadronic stage  might then account for most of the fall off of the
integral $v_{2}$ signal going to forward rapidities.  

Hydrodynamic calculations are able to reproduce the observed behavior.  
This is seen by the open circles in Figure~\ref{fig:fig1} which show the
results of the hybrid hydrodynamic calculations of Hirano {\it et al.} 
(\cite{Hirano:2005xf}, and private communication) that include
dissipative effects of the late hadronic expansion stage. 
Good
agreement is found with the experimental results.
The open crosses in Figure~\ref{fig:fig1} show the results of the AMPT
model for pseudorapidity $\eta = 4$. At this angle the string melting
mechanism that allows for a good reproduction of mid-rapidity results
has been turned off~(\cite{Chen:2004vh}, and private communication).   
Although the comparison is subject to poor matching of pseudorapidity,
there is a suggestion that the longitudinal extent of the produced
medium may be greater than that assumed in current AMPT calculations.

In conclusion, BRAHMS has measured $v_{2}(p_{t},\eta )$ for pions, kaons,
and protons at $\sqrt{s_{NN}}$ = 200~GeV for the Au+Au reaction.  The results 
indicate a longitudinally extended region is produced where
the eccentricity of the created medium and the corresponding pressure
gradients remain remarkably constant.  Hydrodynamic calculations 
with final stage dissipation are found to be in excellent 
agreement with the measured differential $v_{2}$ values.

\end{document}